\documentclass[11pt]{article}
\usepackage{graphicx,amsmath,amsthm,mathrsfs,amssymb}
\newcommand{\ee}{\end{equation}}
\newcommand{\word}[1]{\,\,\mbox{#1}\,\,}
\newcommand{\reff}[1]{(\ref{#1})}
\newcommand{\beq}{\begin{equation}}
\newcommand{\eeq}[1]{\label{#1}\end{equation}}
\newcommand{\beg}{\begin{equation*}}
\newcommand{\eeg}{\end{equation*}}

\newcommand{\eq}{\!=\!}
\newcommand{\p}{\!+\!}
\newcommand{\m}{\!-\!}

\newcommand{\bsplit}{\begin{split}}
\newcommand{\esplit}{\end{split}}

\begin{document}
\begin{titlepage}
\title{Extremal black holes, gravitational entropy \\and nonstationary metric fields  \thanks{based on an essay selected for honorable mention in the 2010 Gravity Research Foundation Essay Competition}}
\author{$^{1,2}$\thanks{aedery@ubishops.ca} Ariel Edery and  $^1$\thanks{bconstantine07@ubishops.ca} Benjamin Constantineau \\\\$^1${\small\it Physics Department, Bishop's University}\\
{\small\it 2600 College Street, Sherbrooke, Qu\'{e}bec, Canada
J1M~0C8}\vspace{0.5em}\\$^2$\thanks{work partly completed at KITP,
Santa Barbara} {\small\it Kavli Institute for Theoretical Physics,
University of California}\\{\small\it Kohn Hall, Santa Barbara,
CA 93106 USA }\vspace{0.8em}}
\date{}
\maketitle

\begin{abstract}
We show that extremal black holes have zero entropy by pointing out a simple fact: they are time-independent throughout the spacetime and correspond to a single classical microstate. We show that non-extremal black holes, including the Schwarzschild black hole, contain a region hidden behind the event horizon where all their Killing vectors are spacelike. This region is nonstationary and the time $t$ labels a continuous set of classical microstates, the phase space $[\,h_{ab}(t), P^{ab}(t)\,]$, where $h_{ab}$ is a three-metric induced on a spacelike hypersurface $\Sigma_t$ and $P^{ab}$ is its momentum conjugate. We determine explicitly the phase space in the interior region of the Schwarzschild black hole. We identify its entropy as a measure of an outside observer's ignorance of the classical microstates in the interior since the parameter $t$ which labels the states lies anywhere between $0$ and $2M$. We provide numerical evidence from recent simulations of gravitational collapse in isotropic coordinates that the entropy of the Schwarzschild black hole stems from the region inside and near the event horizon where the metric fields are nonstationary; the rest of the spacetime, which is static, makes no contribution. Extremal black holes have an event horizon but in contrast to non-extremal black holes, their extended spacetimes do not possess a bifurcate Killing horizon. This is consistent with the fact that extremal black holes are time-independent and therefore have no distinct time-reverse.    
\end{abstract}

\thispagestyle{empty}
\end{titlepage}

\setcounter{page}{1}

\section{Introduction}

It has been over 35 years since Bekenstein's seminal discovery that black holes contain entropy proportional to the area of the event horizon \cite{Bekenstein}. At the classical level, it is well known that black hole mechanics obey laws similar to the ordinary laws of thermodynamics (see \cite{Wald1} for a review). A very useful formula for the entropy of black holes in terms of the Noether charge was later derived in the 90's \cite{Wald2}. Nonetheless, a crucial question has remained unanswered in the area of classical black hole thermodynamics: what is the origin of the entropy at the classical level? More precisely, what are the classical microstates that correspond to the entropy macrostate? This question is directly related to another long-standing issue in the literature which has recently attracted renewed interest: do extremal black holes have zero or non-zero entropy?(see \cite{Randall} for a discussion and references). In 1986, it was proved that black holes obey the weak version of the third law of thermodynamics: the surface gravity $\kappa$ of a black hole cannot be reduced to zero within a finite advanced time \cite{Israel}. We will show that extremal black holes have zero entropy, so that black holes also obey the strong version of the third law.   

Consider a solution to the equations of motion of a Lorentzian field theory which corresponds to a single (nondegenerate) static or stationary field throughout the entire spacetime. Since there is a single field which is time-independent everywhere, there is only one field configuration and this implies that the entropy of the solution is zero. This result applies to any field and the metric (gravitational) field is no exception. A solution to Einstein's field equations which corresponds to a single static or stationary metric field throughout the entire spacetime, has zero entropy since the metric field does not change with time and only a single metric field configuration exists. If so-called static or stationary black holes like Schwarzschild, Reissner-Nordstr\"{o}m or Kerr were actually static or time-independent throughout the entire spacetime, they would certainly have zero entropy.  However, as we will see, except for the extremal cases, they all contain a nonstationary region hidden behind the event horizon. Bekenstein's view, which we adhere to, is that black hole entropy is a measure of an outside observer's inaccessibility of information about its internal configurations \cite{Bekenstein}. In other words, it is a measure of an outside observer's ignorance of the internal configurations hidden behind the event horizon. As far as we know, the possible internal classical configurations have never been clearly identified. In this paper, we fill this gap in the literature. We show that the region behind the event horizon of non-extremal black holes is nonstationary and that the internal classical configurations are points in phase space, the classical microstates $[h_{ab}(t),P^{ab}(t)]$ where $h_{ab}$ is three-metric induced on a spatial hypersurface $\Sigma_t$ and $P^{ab}$ is its momentum conjugate. There is a continuous set of classical microstates parametrized by the time $t$ and an outside observer is ignorant of which classical microstate the black hole interior is in. However, if the phase space in the interior region of a black hole is time-independent, there is only a single classical microstate and the entropy is then clearly zero. Recent numerical simulations of classical gravitational collapse in isotropic coordinates show that classical black hole entropy stems from the region just behind the event horizon where the metric fields are nonstationary. Note that ``nonstationary", ``static" or ``time-independent" are defined here in a coordinate-independent fashion. A nonstationary region is one where all Killing vectors are spacelike; a static region is one that possesses a timelike hypersurface-orthogonal Killing vector and a time-independent region is defined here as a region where the Killing vectors are not all spacelike. In particular, one cannot turn a static or time-independent region into a nonstationary one via a coordinate transformation and vice-versa. 

\section{Zero entropy of extremal RN black hole}

Consider the Reissner-Nordstr\"{o}m (RN) black hole. Its metric is given by $ds^2\eq B(r) dt^2 \m B(r)^{-1} dr^2 \m r^2 \,d\Omega^2$ where $B(r)\eq 1\m 2\,M/r \p Q^2/r^2\eq (1\m M/r)^2 \p (Q^2-M^2)/r^2$ with $Q$ the electric charge and $M$ the (ADM) mass of the black hole. The extremal black hole corresponds to the special case when $Q^2\eq M^2$ and its metric is 
\beq                                 
                       ds^2=  \Big(1- \dfrac{M}{r}\Big)^2 \,dt^2 - \Big(1- \dfrac{M}{r}\Big)^{-2} \,dr^2 - r^2 d\Omega^2 \,.
\eeq{equation}
The above metric is time-independent everywhere. The metric coefficients do not depend on the time coordinate $t$ and never switch sign. In contrast to the Schwarzschild metric, the coordinate $r$ never becomes timelike: it is spacelike both inside and outside the event horizon at $r\eq M$ and is null at the event horizon. The Killing vector $K^{\mu}\equiv (\partial_t)^{\mu} \eq (1,0,0,0)$ has norm $K_{\mu}K^{\mu}\eq (1 \m  M/r)^2$ which is nonnegative. Therefore $K^{\mu}$ is never spacelike: it is timelike (and hypersurface-orthogonal) throughout the spacetime except at the horizon itself where it is null. {\it There is no region where all the Killing vectors for the extremal RN black hole are spacelike.} By definition, there is no region which is nonstationary. 

The metric \reff{equation} has a coordinate singularity at the event horizon $r\eq M$. After performing an ``Eddington-Finkelstein" type coordinate transformation $T = t+r- \frac{M^2}{r-M}+2\,M\,\ln(|r-M|)$ one obtains the metric in the following form:
\beq
ds^2=  \Big(1- \dfrac{M}{r}\Big)^2 \,dT^2 - 2\,dT\,dr - r^2 d\Omega^2 \,.
\eeq{EF}
The metric coefficients are now finite at the event horizon, have no dependence on the time $T$ and never switch sign. The extremal RN black hole is manifestly time-independent everywhere and clearly has zero entropy since it corresponds to a single metric field configuration. In particular, the interior region is time-independent and an outside observer is not ignorant of the internal configuration behind the event horizon. Our result is in agreement with the work of Hawking, Horowitz and Ross \cite{Horowitz} which is based on the Euclideanized solutions. Our approaches are different and complementary. In \cite{Horowitz}, the zero entropy of the extremal RN black hole is attributed to the trivial topology $S^1 \!\times\! R \!\times\! S^2$ of the Euclideanized solution whereas in our case it is attributed to the time-independence everywhere of the Lorentzian solution. We see that black holes obey the strong version of the third law of thermodynamics because extremal black holes have zero surface gravity (temperature of absolute zero) and zero entropy. 

Note that the time-independence of the extremal RN black hole is in agreement with how ordinary classical systems behave at absolute zero. If one considers an isolated classical system at fixed energy $U$ (microcanonical ensemble), one expects that it will be static or time-independent at a temperature $\tau$ of absolute zero. This is certainly the case for simple systems such as the classical ideal gas of $N$ non-interacting particles of mass $m$ ($U\eq3\,N\,\tau/2$,$\,\,H\eq\sum \frac{p_i^2}{2m}$) or a set of N linear harmonic oscillators of frequency $\omega$ and mass $m$ [$U\eq N\tau$,$\,\,H\eq \sum \big( \frac{p_i^2}{2m} + \frac{1}{2}\,m\,\omega^2\,q_i^2\big)]$. At absolute zero, the internal energy $U$ in both cases is zero and from the form of the Hamiltonian $H$, this implies that all momenta ${\bf p}_i$ are zero and that the systems are static. 

The naked singularity RN solution, the case $Q^2 \!> \!M^2$ provides a test case for our approach. We know a priori that this solution has zero entropy since it does not possess an event horizon. This is precisely what we obtain from an analysis similar to the one carried out above for the extremal case. When $Q^2 \!> \!M^2$, the metric coefficient $B(r)\eq (1 \m M/r)^2 \p (Q^2 \m M^2)/r^2$ is positive for all values of $r$ so that the coordinate $r$ is spacelike everywhere. The Killing vector $K^{\mu}$ has norm $K_{\mu}K^{\mu}\eq B(r)$ which is always positive; it is everywhere timelike (and hypersurface-orthogonal) and this implies the spacetime is static everywhere. It therefore has zero entropy because it corresponds to a single metric field configuration. 

\section{Phase space of Schwarzschild interior}

In contrast to the extremal case, non-extremal black holes contain a nonstationary region hidden behind their event horizons. Consider the Schwarzschild case whose metric expressed in standard form is  $ds^2 \eq (1 \m 2\,M/r) dt^2 \m (1 \m 2\,M/r)^{-1} dr^2 \m r^2 d\Omega^2$. The Killing vector $K^{\mu} \equiv (\partial_t)^{\mu} \eq (1,0,0,0)$ has norm $K_{\mu}\,K^{\mu}\eq 1 \m2\,M/r$ and is timelike (and hypersurface-orthogonal) in the exterior region $r\!>\!2\,M$ and null at the event horizon $r\eq 2M$. However, in the interior region ($0 \!<\! r \!<\! 2\,M$) it is spacelike since its norm becomes negative. {\it In the interior region, all the Killing vectors of the Schwarzschild black hole are spacelike.} In that region, the coordinate $r$ becomes timelike, the coordinate $t$ becomes spacelike and the metric fields are nonstationary. In the interior, one can express the metric in the following manifestly nonstationary form \cite{Hawking2}:
\beq
ds^2 \eq \Big(\dfrac{2M}{t} \m 1\Big)^{-1} dt^2 - \Big(\dfrac{2\,M}{t}\m 1\Big) dr^2 \m t^2 d\Omega^2 \,\,; \,\,0 <t<2\,M \,,\, 0<r<\infty \,.
\eeq{time}
To obtain the phase space in the interior region one must foliate the spacetime into a family of spacelike hypersurfaces $\Sigma_t$ at each instant of time \cite{Poisson}. The metric \reff{time} is already in the form $ds^2= N^2\,dt^2 + h_{ab}\,dy^a\,dy^b$ where $N$ is called the lapse function and $h_{ab}$ is the induced metric on the three-dimensional spatial hypersurface $\Sigma_t$. The dynamical phase space is $(h_{ab}, P^{ab})$ where $P^{ab}$ is the momentum conjugate to $h_{ab}$ defined by \cite{Poisson}:
\beq
P^{ab}\equiv \dfrac{\partial}{\partial\dot{h}_{ab}}(\sqrt{-g}\mathcal{L}_G)=\dfrac{\sqrt{-h}}{16\,\pi}(K^{ab} - K\,h^{ab})
\eeq{pab}
where $K_{ab}=\frac{\dot{h}_{ab}}{2\,N}$ is the second fundamental form or extrinsic curvature of the hypersurface $\Sigma$, $h$ is the determinant of $h_{ab}$, $g$ is the determinant of the full (4D) metric $g_{\mu\nu}$ and $\mathcal{L}_G$ is the gravitational Lagrangian (we do not write it out here, see \cite{Poisson}). The lapse function $N$, three-metric $h_{ab}$ and extrinsic curvature $K_{ab}$ in the interior region are given by 
\beq
\begin{split}
&N\eq\Big(\dfrac{2\,M}{t}-1\Big)^{-1/2} \quad;\quad h_{rr}\eq-\big(\dfrac{2M}{t}-1\big) \quad;\quad h_{\theta\theta}\eq -t^2\quad;\quad h_{\phi\phi}\eq -t^2\,\sin^2\theta \\ &K_{rr}\eq \dfrac{M}{t^2}\big(\dfrac{2M}{t}-1\big)^{1/2}\quad;\quad
K_{\theta\theta}\eq -t \,\Big(\dfrac{2M}{t}-1\Big)^{1/2}\quad;\quad K_{\phi\phi}\eq \sin^2\theta K_{\theta\theta}\,.
\end{split}
\eeq{hab2}
Evaluating $P^{ab}$ via \reff{pab} yields
\beq
P^{rr}\eq \dfrac{t\sin\theta}{8\pi}\quad;\quad P^{\theta\theta}\eq \dfrac{-\sin\theta}{16\,\pi}\Big(\dfrac{-M}{t^2} +\dfrac{1}{t}\Big) \quad;\quad P^{\phi\phi}\eq\dfrac{P^{\theta\theta}}{\sin^2\theta}\,.
\eeq{ppab} 
In the interior region, the phase space $[h_{ab}(t), P^{ab}(t)]$ is given by \reff{hab2} and \reff{ppab}. The Hamiltonian constraint ${}^3R \p K_{ab}\,K^{ab}\m K^2\eq 0$ is satisfied in a nontrivial fashion in the interior: 
\beq
{}^3 R \eq \dfrac{-2}{t^2} \,;\,K_{ab}K^{ab}\eq \dfrac{M^2}{t^4}\Big(\dfrac{2\,M}{t}-1\Big)^{-1} \p \dfrac{4M}{t^3} \m\dfrac{2}{t^2} \,; \,\,K^2\eq \dfrac{M^2}{t^4}\Big(\dfrac{2\,M}{t}-1\Big)^{-1} \p \dfrac{4M}{t^3} \m\dfrac{4}{t^2}\,
\eeq{hamiltonian}
where ${}^3R$ is the Ricci scalar constructed from $h_{ab}$ and $K\equiv K_{ab}\,h^{ab}$ is the trace of the extrinsic curvature. In contrast, the Hamiltonian constraint is satisfied in a trivial fashion in the exterior region: ${}^3R \eq0$, $K_{ab}\eq0$ and $K\eq0$. The momentum conjugate $P^{ab}$ calculated via \reff{pab} is zero everywhere in the exterior region. This is expected since the exterior region is completely static.  

The phase space $[h_{ab}(t), P^{ab}(t)]$ in the interior given by \reff{hab2} and \reff{ppab} does not correspond to a single classical microstate but to a continuous set of classical states parametrized or labeled by $t$. The entropy of the Schwarzschild black hole is a measure of an outside observer's ignorance of which classical microstate the black hole interior is in since the parameter $t$ which labels the microstates lies anywhere between $0$ and $2M$. This is in accordance with Bekenstein's view of black hole entropy as``inaccessibility of information about its internal configuration" \cite{Bekenstein}. What we have done here is identified the phase space, the possible internal configurations. If we consider the Schwarzschild black hole as an isolated system with fixed energy $M$ in thermal equilibrium (microcanonical ensemble), the metric fields $h_{ab}(t)$ and their momentum conjugate $P^{ab}(t)$ change with time in the interior region while the entropy macrostate remains basically constant. This is of course what we observe in ordinary classical statistical mechanics: at the microscopic level, the positions and momenta of particles change with time while the macroscopic thermodynamic parameters such as temperature and entropy remain constant. We now present recent numerical evidence that the entropy of the Schwarzschild black hole stems from the interior nonstationary region. 

\subsection{Thermodynamics of gravitational collapse: numerical results} 

The classical gravitational collapse of a spherically symmetric 5D Yang-Mills instanton, in isotropic coordinates, was recently studied numerically \cite{Khlebnikov}. The authors track a thermodynamic function which can be identified with the free energy $F\eq E\m TS$ at late stages of the collapse where $E$ is the $ADM$ mass, $T$ the Hawking temperature and $S$ the entropy. Initially, at the start of the collapse, the instanton is static, the spacetime is nearly Minkowskian and $F\eq E$. At late stages of the collapse, the function $F$ approaches a numerical value close to $E/3$ so that the product $S\,T$ approaches $2E/3$ in agreement with standard black hole thermodynamics \cite{Bekenstein, Hawking1} applied to the 5D Schwarzschild case (i.e. $T \eq \hbar/4\pi r_0$, $S \eq 4\pi^2\,r_0^3/\hbar\,G_5$, $r_0^2 \eq2\,G_5\,E/3\pi$ so that $S\,T =2E/3$. $r_0$ is the gravitational radius and $G_5$ is Newton's constant in 5D). The event horizon in the simulation occurs at $r\eq 0.95$. The $-2E/3$ contribution to the free energy stems entirely from a thin slice in the interior region near the event horizon ($0.95-\epsilon<\!r\!<0.95$) \cite{Khlebnikov} ($\epsilon$ is positive and  approaches zero as the collapse time $t$ approaches $\infty$ where $t$ is time as measured by an asymptotic observer). In other words, the black hole entropy originates from this thin slice in the interior. A plot of the time-derivative of the metric field reveals that it is precisely in this thin slice that the metric field is nonstationary (see Fig. 1 which contains a detailed description)\footnote{ Fig. 1 does not appear in \cite{Khlebnikov}. We therefore coded the differential equations in \cite{Khlebnikov}, reproduced their results, and then plotted Fig. 1 using the data from the run.}. The rest of the spacetime, inside the event horizon ($r\!<0.95-\epsilon$) and outside the event horizon ($r\!>0.95$) is static and makes no contribution to the entropy. {\it The black hole entropy stems from the interior region near the surface of the event horizon where the metric field is nonstationary.} An advantage of using isotropic coordinates is that one can observe the gravitational entropy accumulating directly near the event horizon.  
\begin{figure}
\includegraphics[scale=0.50]{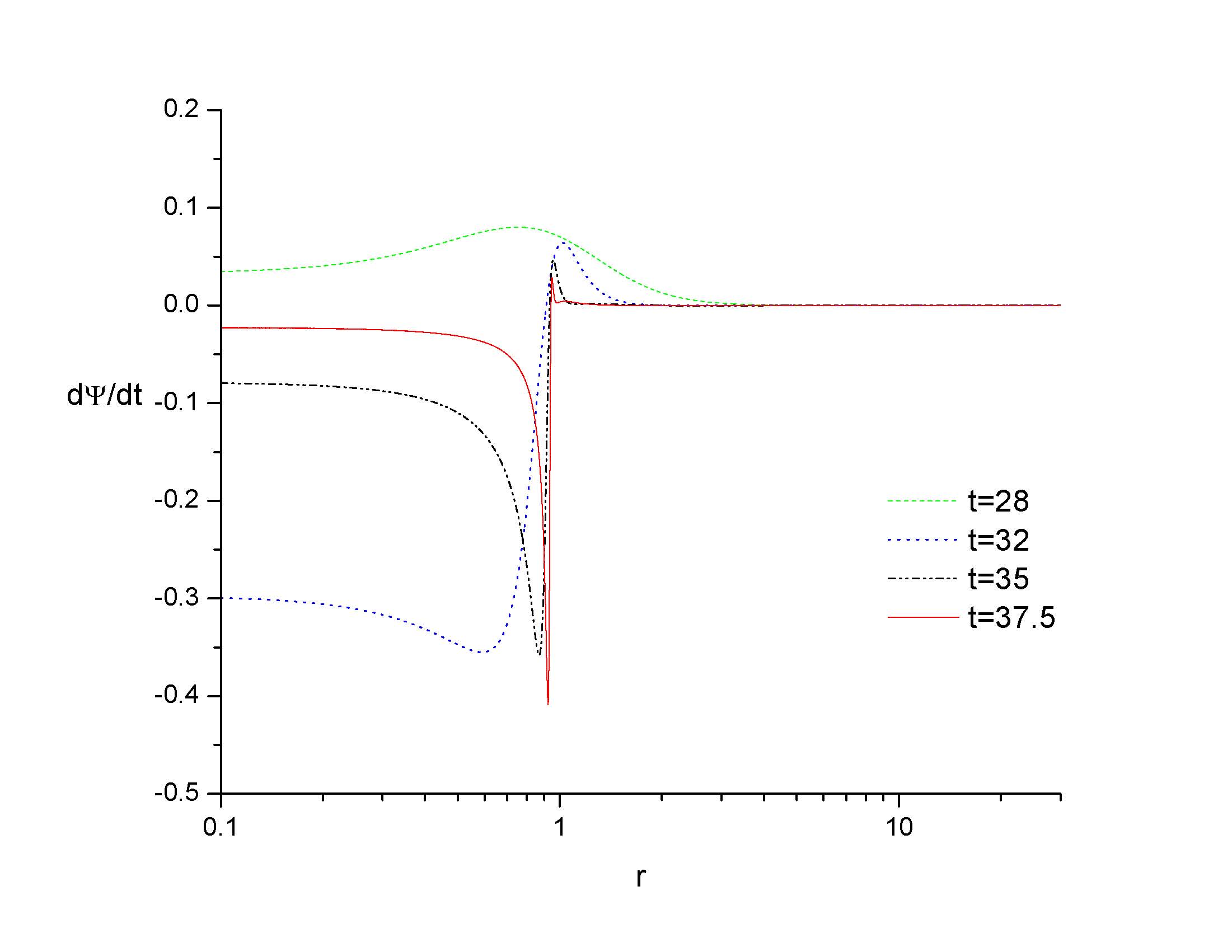}
\caption{The 5D metric in isotropic coordinates is given by $ds^2=N^2(r,t) dt^2 -\Psi^2(r,t)(dr^2 +r^2d\Omega_3^2)$. $d\Psi/dt$ is plotted as a function of $r$ at different times $t$. The radius $r$ and the time $t$ are dimensionless. In the simulation, $r=0.95$ corresponds to the event horizon and the speed of light is set to unity. The plot starts at $t\eq 28$ because the spacetime remains reasonably flat before then. At late stages of the collapse, as time increases from $t=32$ to $t=37.5$, $d\Psi/dt$ decreases significantly inside the event horizon and approaches a value close to zero. Outside the event horizon, it drops abruptly to zero as time increases. However, in the interior, in the vicinity of the event horizon it increases significantly and a large peak is observed. Near thermal equilibrium ($\approx t \eq 37.5$), the metric field $\Psi(r,t)$ is basically static everywhere except on a thin slice in the interior near the event horizon i.e. $0.95-\epsilon< r<0.95$. It is precisely this small region which is responsible for the $-2\,E/3$ contribution to the free energy and hence the black hole entropy. Note that $N$ is not a dynamical variable.} \end{figure}

\section{Extremal Kerr and the non-extremal RN and Kerr black holes } 
As with the Schwarzschild black hole, the non-extremal RN and Kerr black holes contain a nonstationary region hidden behind their event horizon. For both, the nonstationary region is found between the inner and outer horizons. Consider first the non-extremal RN black hole (the case $Q^2 \!< \!M^2$). It has two horizons situated at $r_{\pm}=M\pm\sqrt{M^2-Q^2}$ where $r_{\p}$ and $r_{\m}$ are the outer and inner horizons respectively.  It is convenient to express the metric coefficient $B(r)$ (previously defined above) in terms of $r_{\p}$ and $r_{\m}$ i.e. $B(r)\eq (r-r_{\p})(r-r_{\m})/r^2$. The Killing vector $K^{\mu} \equiv (\partial_t)^{\mu} \eq (1,0,0,0)$ has norm $K_{\mu}\,K^{\mu}\eq B(r)$. When $r\!>\! r_{\p}$ or $r\!<\!r_{\m}$ , $B(r)$ is positive, the coordinate $r$ is spacelike and the Killing vector is timelike. At the horizons, $B(r)\eq0$,  $r$ is null and the Killing vector is null. However, in the region between the two horizons, $r_{\m}\!<\!r\!<\! r_{\p}$ , $B(r)$ is negative, the coordinate $r$ becomes timelike and the Killing vector $K^{\mu}$ is spacelike. This region is nonstationary since all the Killing vectors in that region are spacelike. The non-extremal black hole has a non-zero entropy because the phase space has a time-dependence in the region in between the two horizons and more than one classical microstate is hidden behind the event horizon. We do not explicitly evaluate the phase space since the procedure is similar to what was illustrated for the Schwarzschild case. We now turn to the Kerr black hole.     

The Kerr metric in standard Boyer-Lindquist coordinates is given by $ds^2\eq \tfrac{\rho^2 \,\Delta}{\Sigma}\, dt^2$\\
$ - \tfrac{\Sigma\sin^2\theta}{\rho^2} \,(d\phi-\omega\,dt)^2$ $ -\tfrac{\rho^2}{\Delta}\,dr^2 -\rho^2\,d\theta^2$ \cite{Wald3,Carroll,Poisson} where $a$ is the rotational parameter, $M$ the (ADM) mass, $\rho^2\equiv r^2+a^2\cos^2\theta$, $\Delta\equiv r^2-2Mr +a^2$, $\Sigma\equiv (r^2+a^2)^2-a^2\,\Delta\,\sin^2\theta$ and $\omega\!\equiv \! \tfrac{-g_{t\phi}}{g_{\phi\phi}}\eq \tfrac{2Mar}{\Sigma}$. Solutions to $\Delta\eq 0$ yield the two horizons, $r_{\pm}=M \pm\sqrt{M^2-a^2}$ where $r_{\p}$ and $r_{\m}$ are the outer and inner horizons respectively.  The Kerr metric has two Killing vectors, $K^{\mu}\eq (\partial_t)^{\mu}\eq (1,0,0,0)$ and  $R^{\mu}\eq (\partial_{\phi})^{\mu}\eq (0,0,0,1)$ and one can construct a Killing vector $\chi^{\mu}$ which is a linear combination of the two: $\chi^{\mu}\eq  K^{\mu} +\omega_0\,R^{\mu}$ where $\omega_0$ is a constant. After some algebra, one obtains the following formula for the norm of $\chi^{\mu}$:
\beq
 \chi^{\mu}\,\chi_{\mu}=\dfrac{\rho^2 \Delta}{\Sigma} - \dfrac{\Sigma\,\sin^2 \theta}{\rho^2} (\omega_0 -\omega)^2
 \eeq{KillingVector}
where $\omega$ is defined above. The norm is the sum of two terms. With $\rho^2$ and $\Sigma$ being both positive (except at the ring singularity where they are both zero), the second term is negative or zero. The sign of the first term depends on the sign of $\Delta$. Consider the non-extremal case. $\Delta$ can be expressed as $(r-r_{\p})(r-r_{\m})$ and it is positive for $r\!>\! r_{\p}$ or $r\!<\! r_{\m}$ and zero at $r\eq r_{\p}$ and $r\eq r_{\m}$. However, $\Delta$ is negative in the region $r_{\m}\!<\!r\!<\!r_{\p}\,\,$, the region $\Re$ in between the two horizons. This implies that the region $\Re$ is nonstationary since any linear combination of the two Killing vectors is spacelike in that region. The non-extremal Kerr black hole has a non-zero entropy for the same reasons as the Schwarzschild and non-extremal RN black hole. 

For the extremal Kerr black hole ($a\eq M$), $\Delta$ is equal to $(r-a)^2$ which is positive everywhere except at the event horizon $r \eq a$ where it is zero. The first term in \reff{KillingVector} is therefore positive or zero. Let $\omega_0$ be equal to $\omega$ evaluated at $r\eq r_0$. Physically, this corresponds to the angular velocity of a zero angular momentum observer (ZAMO) situated at $r\eq r_0$ \cite{Poisson}. At this location, the second term in \reff{KillingVector} is zero and with the first term being positive or zero, this implies the Killing vector $\chi^{\mu}$ at $r\eq r_0$ is timelike (or null). At each point, one can construct a different timelike Killing vector except at the horizon where it would be null. For the extremal Kerr black hole, there is no point in the entire spacetime where all the Killing vectors are spacelike and no region is nonstationary. Local stationary observers (the ZAMOs) exist at every point. As we will see, there is no time evolution of the phase space $[h_{ab},P^{ab}]$: the induced three-metric $h_{ab}$ and its momentum conjugate $P^{ab}$ are both time-independent everywhere. The extremal Kerr black hole has zero entropy because it has only a single classical microstate. To see this, we write the metric in the general $3+1$ decomposition \cite{Poisson} and extract the lapse function $N$, the shift vector $N^a$ and the induced metric $h_{ab}$ i.e.
\beq
\begin{split}
ds^2&=N^2\,dt^2 + h_{ab}\,(dy^a +N^a \,dt)(dy^b +N^b\,dt)\\
&=N^2\,dt^2 +h_{\phi\phi}(d\phi +N^{\phi}\,dt)(d\phi +N^{\phi}\,dt) +h_{rr}\,dr^2 +h_{\theta\theta}\,d\theta^2\\
&=\dfrac{\rho^2 \,\Delta}{\Sigma}\, dt^2 - \dfrac{\Sigma\sin^2\theta}{\rho^2} \,(d\phi-\omega\,dt)^2 -\dfrac{\rho^2}{\Delta}\,dr^2 -\rho^2\,d\theta^2\\
&\word{where}\\
&N^2= \dfrac{\rho^2 \,\Delta}{\Sigma}, h_{\phi\phi}=- \dfrac{\Sigma\sin^2\theta}{\rho^2}, h_{rr} \eq -\dfrac{\rho^2}{\Delta}, 
h_{\theta\theta}\eq -\rho^2 \word{and} N^{\phi} = -\,\omega\,.
\end{split}
\eeq{NP}
For the extremal black hole, $\Delta$ is nonnegative everywhere. $N^2$ is therefore positive or zero and this implies the equation for $N^2$ is valid everywhere since $N^2$ is never negative. Since $\rho^2$, $\Sigma$ and $\Delta$ depend only on $r$ and $\theta$, the lapse function $N$ also depends only on $r$ and $\theta$ and is time-independent. The three-metric coefficients, $h_{\phi\phi}, h_{rr}$ and $h_{\theta\theta}$ do not switch sign (they are negative or zero), depend only on $r$ and $\theta$ and are time-independent: $\dot{h}_{ab}$ is zero everywhere. Same thing for the shift $N^{\phi}$: it is time-independent and depends on $r$ and $\theta$ only. We now show that $P^{ab}$, the momentum conjugate to $h_{ab}$, is also independent of time. We first evaluate the extrinsic curvature $K_{ab}$, which is defined by \cite{Poisson}
\beq
\begin{split}
&K_{ab}\equiv\dfrac{1}{2N} (\dot{h}_{ab}-\nabla_b\, N_a-\nabla_a\, N_b)\,. \word{The non-zero components are}\\
&K_{\phi\theta}=K_{\theta\phi}=\dfrac{1}{2N}(-\partial_{\theta} N_{\phi} +\dfrac{1}{2} \,N^{\phi}\partial_{\theta} h_{\phi\phi})\word{and} K_{\phi\,r}=K_{r\,\phi}=\dfrac{1}{2N}(-\partial_{r} N_{\phi} +\dfrac{1}{2} \,N^{\phi}\partial_{r}h_{\phi\phi})\\
&\!\!\word{where} N_{\phi}=h_{\phi\phi}\,N^{\phi} = \dfrac{\Sigma\,\omega\,\sin^2\theta}{\rho^2}.\word{We also have} K\equiv h^{ab}\,K_{ab} =0\,.
\end{split}
\eeq{gfd}
$P^{ab}$, defined in \reff{pab}, is given by 
\beq
\begin{split}
&P^{ab}=\dfrac{\sqrt{-h}}{16\pi}\,(K^{ab} - K\,h^{ab})= \dfrac{\sqrt{-h}}{16\pi}\,K^{ab}\,.
\word{The non-zero components are}\\
&P^{\phi\theta}=P^{\theta\phi}= \dfrac{\sqrt{-h}}{16\pi}\,K^{\phi\theta} \word{and}
P^{\phi\,r}=P^{r\,\phi}= \dfrac{\sqrt{-h}}{16\pi}\,K^{\phi\,r}\\
&\word{where} K^{\phi\theta} = \dfrac{K_{\phi\theta}}{h_{\phi\phi}h_{\theta\theta}} \word{and} K^{\phi\,r} = \dfrac{K_{\phi\,r}}{h_{\phi\phi}\,h_{r\,r}} \,.
\end{split}
\eeq{pab2}
It is clear from the above construction that $P^{ab}$ has a dependence only on $r$ and $\theta$ and is time-independent. There is only a single classical microstate $[h_{ab}(r,\theta), P^{ab}(r,\theta)]$. There is no time evolution and hence no continuous set of classical states as in the interior of Schwarzschild. In particular, an outside observer is not ignorant of the classical state found in the region inside the event horizon.      

The above arguments do not hold for the non-extremal Kerr black hole since $\Delta$ is negative in the region $\Re$ in between the two horizons and this implies $N^2$ is negative, which is not valid. The region $\Re$ is nonstationary and the correct $3\p1$ decomposition in that region has $N^2$ positive but time-dependent. To obtain the phase space in $\Re$ the procedure is similar to that employed for the interior of the Schwarzschild black hole and we will not explicitly carry it out here. Briefly, one expresses the metric in $\Re$ in nonstationary form and extracts the lapse function $N$, the three-metric $h_{ab}$, the shift vector $N^{a}$ and from them one calculates $P^{ab}$. The phase space in $\Re$, as expected, is time-dependent. There is more than one classical microstate hidden from an outside observer and the entropy is non-zero.   
           
\section{Time-reversal symmetry and bifurcate Killing horizon}                  

Einstein's field equations are time-reversal invariant so that maximally extended spacetimes include not only the black hole solution but also its associated ``time-reverse". Penrose diagrams of maximally extended spacetimes should therefore contain distinct time-reverse patches associated with the nonstationary region of non-extremal black holes. In contrast, no distinct time-reverse should exist for the extremal black hole since they are time-independent everywhere. This is precisely what we observe. We are familiar with the extended Schwarzschild spacetime (Fig. 2a) where the white hole region III is the time-reverse of the black hole region II (in Kruskal coordinates, region II and III are represented by  $0 \!<\!T^2 \m R^2 \!<\! 1$ with $T\!>\!0$ in region II and $T\!<\!0$ in region III \cite{Wald3,Carroll}. The time-reverse of region II yields region III and vice-versa). The black hole region II has a distinct time-reverse because it is nonstationary. Similarly, for the non-extremal RN extended spacetime (Fig. 2b), region C is the time-reverse of the black hole region B. Regions B and C both represent patches in between the two horizons but they are distinct. In region B, a particle must move in a direction of decreasing r, toward the inner horizon $r_{\m}$ whereas in region C it must move in a direction of increasing r toward the outer horizon $r_{\p}$ \cite{Poisson,Carroll}. Again, region B has a distinct time-reverse because the region in between the two horizons is nonstationary. In contrast, note that the extremal RN extended spacetime (Fig. 2c) does not contain a distinct time-reverse patch but only copies of the same patch (patch A). The extremal RN black hole has no time-reverse (except itself) which is equivalent to stating that it is time-independent everywhere. 

The extended spacetime for the non-extremal cases ($\kappa\ne0$) possess a bifurcate Killing horizon whereas the extremal cases ($\kappa\eq0$) do not \cite{Wald1}. In the Schwarzschild case (Fig. 2a), the future event horizon, which separates regions I and II is distinct from the past event horizon which separates the time-reverse region III and region I; the two intersect at the bifurcation two-sphere. In the extremal RN case, there is no distinct time-reverse region and no second (distinct) event horizon to intersect with. The time-independence of the extremal black hole, its zero surface gravity, its zero entropy and the absence of a bifurcate Killing horizon in the extended spacetime are all related properties that distinguish the extremal from the non-extremal case.  Finally, the naked singularity RN solution (Fig. 2d) is composed of only one region and has no time-reverse patch. It is static everywhere and has zero entropy like the extremal case. This is expected since it has no event horizon. 
          
\section{Conclusion}

In this work, we adhere to Bekenstein's view of black hole entropy as a measure of an outside observer's ignorance of the internal configurations hidden behind the event horizon \cite{Bekenstein}. As far as we know, the question of what are the internal classical configurations, does not seem to have been addressed. We identify the internal configurations as points in phase space, the classical microstates $[h_{ab},P^{ab}]$. We found that extremal black holes have zero entropy because they possess a single classical microstate. This is in agreement with the zero entropy result obtained in \cite{Horowitz} using a different approach. Extremal black holes have a temperature of absolute zero (surface gravity $\kappa\eq 0$) and zero entropy so that black holes obey the strong version of the third law of thermodynamics. The surface gravity $\kappa$ of a black hole cannot be reduced to zero within a finite advanced time \cite{Israel} and this implies that extremal black holes cannot be formed via gravitational collapse or via any finite number of processes. It was then argued in \cite{Horowitz} that an extremal black hole can never become a non-extremal black hole and vice-versa by any quantum or classical process. Extremal black holes are therefore best viewed as $eternal$ black holes. It is convenient to separate extremal black holes into two categories: the extremal limit of a non-extremal black hole and an (eternal) extremal black hole. The two are discontinuous: the extremal limit case has $\kappa\ne0$ and a bifurcate Killing horizon whereas the eternal case has $\kappa \eq 0$ and no bifurcate Killing horizon. It has been argued that the region in between the two horizons does not disappear in the extremal limit \cite{Randall} and from our work, this implies a non-zero entropy. This could explain why some authors obtain a non-zero entropy for what they refer to as extremal black holes (see \cite{Randall} for a full discussion).

There are two conditions for the existence of a non-zero black hole entropy: an event horizon to hide the internal configurations and more than one internal configuration to hide. Extremal black holes have an event horizon but because the phase space is time-independent, they do not hide more than one internal configuration. They do not possess a distinct time-reverse and this shows up in the maximally extended spacetime as an absence of a bifurcate Killing horizon. The condition for a non-zero entropy, at least for the black holes considered here, is the existence of a bifurcate Killing horizon in the extended spacetime. This single criteria satisfies both conditions discussed above. In this context, it is worth noting that Wald's general formula for the entropy of a black hole \cite{Wald2, Wald3} is evaluated at the bifurcation two-sphere not the event horizon. 
   
\begin{figure}[t]
\includegraphics[scale=0.50]{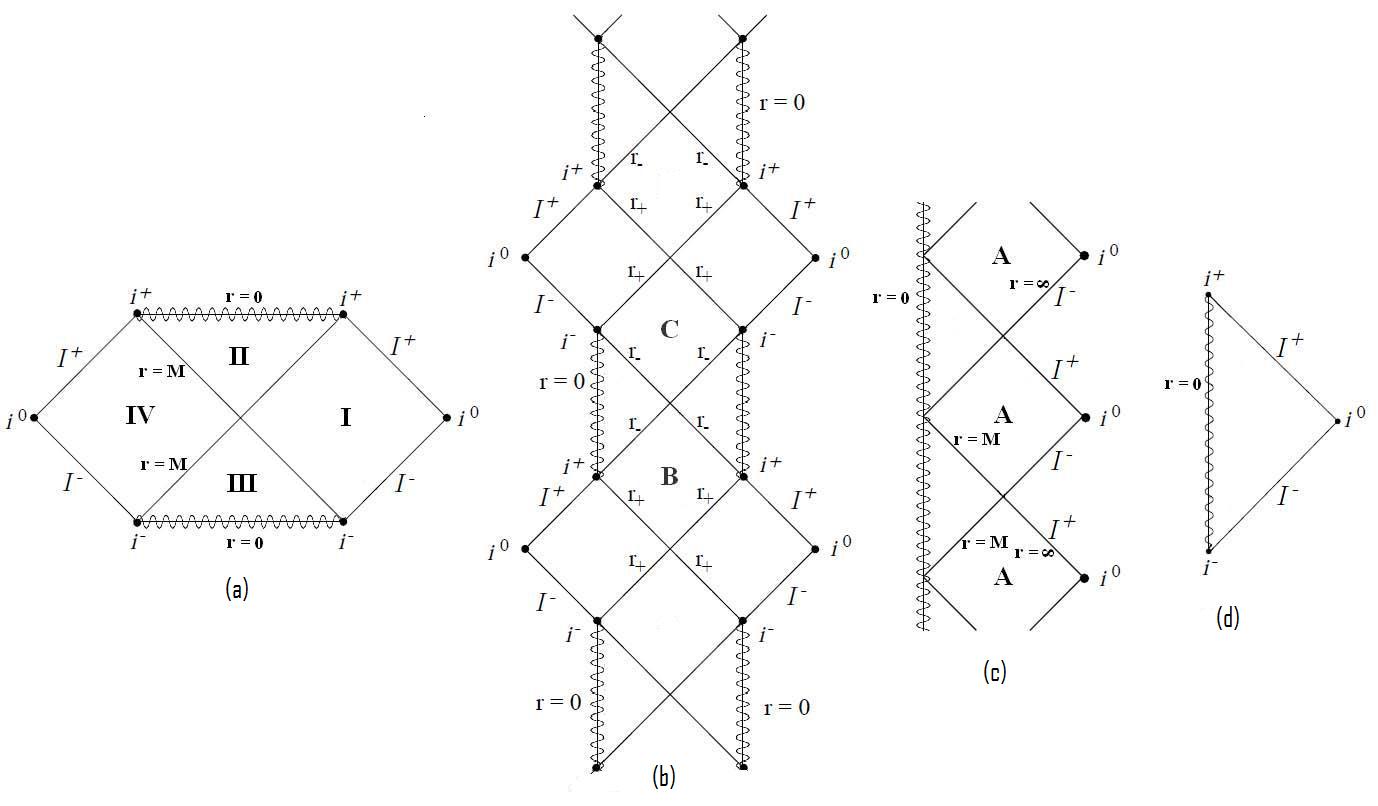}
\caption{Maximally extended spacetimes for (a) Schwarzschild (b) non-extremal RN (c) extremal RN and (d) RN naked singularity.}
\end{figure}

\section*{Acknowledgments}
A.E. acknowledges support from a discovery grant of the National
Science and Engineering Research Council of Canada (NSERC) and from
a Bishop's Senate Research Grant. A.E. would like to thank the organizers 
of the TPI's 50th anniversary celebration at the University of Alberta where 
a talk on this subject was presented. We thank Valerio Faraoni and Manu Paranjape 
for discussions. This research was supported in part by the National Science Foundation 
under Grant No. PHY05-51164.

\end{document}